\documentstyle[aps,pra,multicol,epsf]{revtex}
\begin{document}
\draft
\title
{Dissipative Dynamics in a Quantum Register}
\author {P. Zanardii $^{1,2}$}
\address{$^1$ ISI Foundation, Villa Gualino\\
Viale Settimio Severo 65 -101133 Torino, Italy\\
 $^2$ Unit\`a INFM, Politecnico di Torino,\\
Corso Duca degli Abruzzi 24, I-10129 Torino, Italy
} 
\maketitle
 \begin{abstract}
{ A  model for a quantum register dissipatively coupled with a bosonic thermal bath
 is studied. The register consists  of $N$ qubits (i.e.  
spin $\frac{1}{2}$ degrees of freedom), the bath is described
by $N_b$ bosonic modes. The register-bath coupling is chosen
in such a way that the total number of excitations is conserved.
The Hilbert space splits allowing the study of the dynamics separately
in each sector. Assuming  that the coupling with the bath
is the same for all  qubits, the
excitation sectors have  a further decomposition
according the irreducible representations of the $su(2)$ 
spin algebra. 
The stability against environment-generated noise of 
the information encoded in a  quantum
state of the register
depends on its $su(2)$ symmetry content.
At zero temperature we find that states belonging to the vacuum symmetry sector
have for long time vanishing fidelity, whereas 
 each lowest
 spin vector is decoupled from the bath
and therefore is decoherence free. 
 Numerical results are shown in the one-excitation space in the case
qubit-dependent bath-system coupling.}
\end{abstract}
\begin{multicols}{2}[]
\narrowtext
\section
{introduction}
The unavoidable interaction that each real-world system has with its environment
is one of the major limitations to pratical realization of a {\sl quantum computer} \cite{QC}.
Indeed the outstanding potential capabilities of such a device rely heavily on the possibility
of maintaining the quantum coherence in the system, and one of the typical effects
of the coupling with the environment is to destroy phase relations between 
quantum states appearing in a linear superposition.
This latter phenomenon is referred to as {\sl decoherence} \cite{ZUDI}: it can take place
also when there is no system-environment energy exchange at all. 
To overcome this difficulty, in the last few years there has been a growing interest
in the so-called {\sl error correction} schemes \cite{ERROR}, in which, by means of suitable
encondings and measurement protocols, one is able to disentangle the system
from the environment in order  to recover uncorrupted information.
Another possible approach, pointed out in \cite{ZARA}, is to make use of symmetry
to protect the information stored in a quantum state against environment-induced noise.
In this case one, rather than to design states that can be easily corrected,
looks for states that cannot easily  be disturbed.
The  simplest system for which dynamical symmetry provides us these "safe" states
is a collection of $N$ qubits coupled all in the same way with the environment.
Such a system can be thought of as quantum register $S,$ analogous
to those characteristic of classical computation.
In this paper we study a model Hamiltonian describing the exchange of elementary
{\sl quanta} between the register and the environment, modelled by a bosonic bath $B$.
The coupling with the bath is realized in term of the off-diagonal 
generators of a $su(2)$ dynamical algebra \cite{SOBI}.
The marginal dynamics of $S$ is dissipative (i.e. the register energy is not conserved).
The global Hilbert space
decomposes in (dynamically) invariant sectors, characterized by their $su(2)$ symmetry content
as well as their  number of excitations, as it will be specified later.
It should be  emphasized that the focus 
of this paper is  on the role played by dynamical-algebraic
structures in providing  collective states of the register 
intrinsecally stable against environment-induced decoherence.
The adopted physical model is, in a sense,  {\sl generic} and it is not
aimed to describe a  specific physical implementation  of a Quantum Computer
(as done instead, for example in \cite{GAR})  but a broad class of open quantum systems
that could eventually turn out to be relevant for quantum data processing
applications. \\  
In sect. II, after recalling the  fundamentals of open quantum systems,
 the model is introduced and its general feature briefly discussed.
In sect. III the associated Hilbert space structure is analyzed.
In sect. IV the one-excitation subspace is studied,
 analytical as well as
numerical results are presented for the fidelity and entropy.
 Sect. V contains some preliminary numerical results in the case
of a qubit depending coupling with the bath.
Section VI contains a number of conclusive remarks and perspectives
\section{the model}%%%%%%%%%%%%%%%%%%%%%%%%%%%%%%%%%%%%%%%%%%%%%%%%%%%555
Before introducing  our model  we begin by
 briefly recalling  a few basic facts about open quantum systems.\\
Let  ${\cal H}_s,\,{\cal H}_b$ denote respectively the system and the environment
 Hilbert
spaces. We assume  ${\cal H}_b$ to be much larger than ${\cal H}_s.$
The total Hilbert space is given by the tensor product ${\cal H}={\cal H}_s\otimes{\cal H}_b.$
A  {\sl state} over ${\cal H}_\alpha$ ($\alpha=s,\,b$) is a hermitean non negative operator $\rho_\alpha$
of $\mbox{End}({\cal H}_\alpha)$ with 
$\mbox{tr}^\alpha
(\rho_\alpha)=1.$
The manifold of the state over ${\cal H}_\alpha$ will be denoted by ${\cal S}_\alpha.$
The elements of ${\cal S}_\alpha$ that are also projectors  
($\rho^2=\rho$) provide the pure states. The set ${\cal S}_\alpha^P$  of pure states
 generates ${\cal S}_\alpha$
as its convex hull, furthermore it is in a one-to-one correspondence with ${\cal H}_\alpha.$
According to quantum mechanics, time evolution of the overall (closed) system is unitary, therefore
if the initial state has the separable form $\rho(0)=\rho_s\otimes\rho_b,\,(\rho_\alpha\in {\cal S}_\alpha)$
 then for any $t\ge 0,$ 
the marginal  (Liouvillian) evolution on ${\cal S}_s$ (open) is given by
\begin{equation}
{\cal E}^{\rho_b}_t\colon {\cal S}_s\rightarrow {\cal S}_s
\colon
\rho_s\rightarrow\mbox{tr}^b\,(U_t\,\rho(0)\,U^\dagger_t),
\end{equation}
where $\mbox{tr}^b$ denotes the partial trace over ${\cal H}_b.$
The {\sl superoperators} $\{{\cal E}^{\rho_b}_t\}_{t\ge 0}$
are trace-preserving completely positive maps \cite{DAV}, that are the most general
description of the evolution of an open quantum system.
 ${\cal S}_\alpha^P$ is  not invariant under the action of $\{{\cal E}^{\rho_b}_t\}_{t\ge 0},$
{\sl typically} an initial pure state of the subsystem becomes mixed in a very short time scale
depending on the strength of the interaction.
This state  can either eventually get pure again or not, but in any case an irreversible
loss of the information stored in the initial preparation has occurred.
It is important to notice that this mechanism is active
even when there is no energy-exchange  between the subsystems (i.e. the subystem Hamiltonians
are constants of motion) at finite as well as at zero temperature.
When an energy-exchange occurs we call the resulting dynamics {\sl dissipative.}
\\
We introduce now the model.
The Hamiltonian is given by $H=H_s+H_b+H_I$ where
\begin{eqnarray}
H_s &=& \sum_{i=1}^{N} \epsilon_i\, \sigma_i^z, \\ \nonumber
H_b &=& \sum_{k=1}^{N_b} \omega_k\,(b_k^\dagger b_k+1/2), \\ \nonumber
H_I &=& \sum_{i=1}^{N} \sum_{k=1}^{N_b} ( g_{ki} b_k^\dagger \sigma^-_i + \mbox{h.c.}),
\end{eqnarray}
here the $\sigma_i^\alpha$'s are the spin $1/2$ Pauli operators ($i$ is the qubit index),
and the $b_k$'s bosonic operators.
$H_s$ ($H_b$) is the  Hamiltonian of the register (bath), $H_I$ the register-bath interaction.
This model is closely related to the one 
 known in the literature as the Dicke maser model \cite{HELI}.
The latter, for generic
$N,\,N_b,$  is not solvable and has a non-trivial ground-state phase diagram.
In order to shed some light on the physics of this system we write down  the equation of the motion
for the Heisenberg operators $O(t)\equiv U(t)^\dagger\,O\,U(t).$
To simplify the expressions, it turns useful to perform the (unitary) transformation 
$\sigma_j^{\pm}\mapsto \sigma_j^{\pm}\exp(\pm\, i\epsilon_i\,t),\,b_k\mapsto b_k\exp(-i\,\omega_k\,t)$,
whereby the Heisenberg equations then read
\begin{eqnarray}
i\,\frac{\partial \sigma_i^+}{\partial t} &=& 2\,\sum_k g_{ki}(t) b_k^\dagger\,\sigma_i^z,\\ \nonumber
i\,\frac{\partial \sigma_i^z}{\partial t}\ &=& -\sum_k (g_{ki}(t) b_k^\dagger\,\sigma^-_i-\mbox{h.c.}),\\ \nonumber
i\,\frac{\partial b_k}{\partial t} &=& -\sum_j g_{kj}(t) \sigma_j^-.
\end{eqnarray}
where $g_{ki}(t)\equiv g_{ki}\exp[i\,(\omega_k-\epsilon_i)\,t]$.
By a formal integration of the field equation and the substitution of the result into the 
spin equations one obtains
\begin{eqnarray}
i\,\frac{\partial \sigma_i^+}{\partial t} &=&
\phi_i^+(t)-2\,i\sum_j\int_0^t d\tau K_{ij}(t,\tau)\sigma_j^+(\tau)\,\sigma_i^z(t),
\\ \nonumber
i\,\frac{\partial \sigma_i^z}{\partial t} &=&- \phi_i^z(t)-\sum_j \int_0^t d\tau
K_{ij}(t,\tau)\sigma_j^+(\tau)\sigma_i^-(t)-\mbox{h.c}.
\end{eqnarray}
Here
\begin{eqnarray}
K_{ij}(t,t^\prime) &=& \sum_k g_{ki}(t)\,g_{kj}^*(t^\prime)
\\ \nonumber
\phi_i^+(t) &=& 2\,\sum_k g_{ki}(t)\, b_k^\dagger(0) \sigma_i^z(t),\\ \nonumber 
\phi_i^z(t) &=& \sum_k g_{ki}(t)\,b_k^\dagger(0)\sigma_i^-(t).
\label{eqmot}
\end{eqnarray}
This coupled system of non-linear integro-differential equations describes
the dynamics of the spin subsystem in closed form.
By means of the intermediation of the bath bosons each spin gets interacting
with all the others {it via} a sort of time-retarded Heisenberg
coupling.
The information about the bath (dynamics as well as
preparation) is contained in the kernels $K_{ij}$,
and in the operators $\{\phi_i^\alpha\}.$
If $\epsilon_j=\epsilon,\,(j=1,\ldots,N)$ and the bath-spin coupling is  the same for all the spins,
one has
\begin{equation}
K_{ij}(t,t^\prime)=\sum_k |g_{k}|^2 \exp[-i(\epsilon-\omega_k)\,(t-t^\prime)].
\end{equation}
Under rather general assumptions  this kernel is strongly peaked at $t=t^\prime$,
If one has $K_{ij}(t,t^\prime)\sim\delta(t-t^\prime)$ then
  (\ref{eqmot}) become a system of coupled non-linear differential equations. 
Despite  this strong simplification, also in this case the solution, due to non-linearity,  remains difficult
and one has to resort to numerical techniques.\\
An alternative approach  based on symmetry considerations,
 will be introduced in the next section.
\section{Hilbert Space Structure}%% %%%%%%%%%%%%%%%%%%%%%%%%%%%%%%%%%%%%%%%%%%%%%%%%%%%%%%%
The Hilbert space is given by the tensor product ${\cal H}={\cal H}_s^{\otimes\,N}\otimes
{\cal H}_b^{\otimes\,N_b}$, being ${\cal H}_s$ (${\cal H}_b$) the two (infinite) dimensional single spin (boson)
space.
The coupling of the spin system with the bosonic bath is described 
by the hamiltonian
 $H_I$ is such that the raising (lowering)
of one spin state is associated to the destruction (creation) of one boson. 
From this follows that the  system admits the  constant of motion
\begin{eqnarray}
{\cal I}=\sum_{i=1}^{N}\sigma^z_i+\sum_{k=1}^{N_b} n_k+N/2.
\end{eqnarray}
The eigenvalues of ${\cal I}$ give the number of elementary (spin as well as bosonic)
excitations over the reference state
$|0\rangle\equiv |0\rangle_s\otimes |0\rangle_b$. 
The latter  is a lowest weight vector for the spin as well
as for the boson algebra:
$\sigma^-_\alpha\;|0\rangle=b_k\;|0\rangle=0,\;\forall \alpha,k.$
Its energy is set equal to zero.
The Hilbert space splits into invariant eigen-spaces of $\cal I$, 
${\cal H}=\oplus_{I}{\cal H}_{I};$ an elementary combinatorial  argument shows that
 the dimension
of the $I$-excitations space ${\cal H}_I,\;(I\in{\bf {N}})$ is given by
\begin{eqnarray}
d_I=\sum_{l=0}^{{ {min}}(N,I)}\pmatrix{N\cr l}
\pmatrix{I-l+N_b-1\cr N_b-1}.\label{Dim}
\end{eqnarray}
If $N=N_b=1,$ one has $d_0=1,\,d_I=2,\,(I\ge 1),$
the model reduces to the exactly solvable Jaynes-Cummings model of quantum
optics \cite{JC}.
It is worth  noticing that the general spin-boson model considered in the literature
on quantum dissipation, [usually addressed in the framework  of the Feynmann-Vernon
influence functional (see \cite{LEG} for a review)], due to the presence 
of terms $b_k\,\sigma^-_i,\,b^\dagger_k\,\sigma^+_i,$ (neglected here in view of the
rotating wave approximation)
{\sl does not} conserve ${\cal I},$
spoiling the associated dynamical decomposition of the Hilbert space,
on which  our subsequent numerical analysis relies. 
Nevertheless, since in this papers we are interested
only in the role played by collective effects in stabilizing  a quantum state,
 this  restrictions
does not result  in any severe loss of generality.
\\
A basis for ${\cal H}_I$ is given by $|\psi^{(n)}_{\alpha,k}\rangle=|\alpha\rangle\otimes|k\rangle,$
where
\begin{equation}
|\alpha\rangle\equiv\prod_{j=1}^n\sigma_{\alpha_j}^+\,|0\rangle_s,\qquad 
|k\rangle\equiv \prod_{j=1}^{I-n}b_{k_j}^\dagger\, |0\rangle_b.
\label{Nbase}
\end{equation}
Where
$ n=1,\ldots,{min}(I,N),\; \alpha\in {\cal C}(N,n),\; k\in {\cal C}^\prime(N_b,N-n),
$
${\cal C}(n,k)$ (${\cal C}^\prime(n,k)$) denoting the set of the combinations without
 (with) repetitions of $n$ objects $k$ by $k$.
Following the general scheme of \cite{ZARA}
  we specialize hereafter the model assuming  the parameters $\{\epsilon_i\},\;\{g_{ki}\},\;(i=1,\ldots,N),$
independent of the qubit replica index $i.$
The first assumptions follows simply from the fact that qubits are {\sl replicas} of the same system.
The independence of the coupling constants on the qubit index is analogous to the so-called Dicke limit
of quantum optics \cite{HELI}; it holds - for example -  when the typical bath wave-lengths coupled with the register
are much greater than the distances between the qubits.
In this latter case the environment is no able to probe the internal structure of the register:
as long as the dynamics is concerned it has  an effective point-like topology.
The common value $\epsilon$ of the qubit "magnetic" fields $\{\epsilon_i\}$ will be chosen as the unit
of the energy scale; analogously the 
"sound velocity" of the boson will be set equal to one, so that
their   dispersion relation reads $\omega_k= k,\;(k=2\,\pi\,n/N_b,\;n=1,\ldots,N_b).$
The Hamiltonian can then be written as 
\begin{equation}
 H= \epsilon \,S^z+ B\,S^+ +S^-\,B^\dagger+H_b,
\label{Hamil}
\end{equation}
 being $S^\alpha=\sum_{j=1}^{N}  S_j^\alpha,\,(\alpha=z,\pm)$ global
spin operators, spanning a Lie Algebra $su(2),$ and $B=\sum_k g_k b_k$.\\
The fact that in (\ref{Hamil}) the spins  appear only
trough the $S^\alpha$'s means the all the qubits are treated symmetrically:  
 the dynamics allows only for coherent excitations of the computational (spin) degrees of freedom.
This is, from the algebraic point of view, a very strong constraint:
the {\sl dynamics gets invariant under the action
of the symmetryc group ${\cal S}_N$ of the qubit index permutations.}
 This provides
us one more constant of motion.
Indeed from (\ref{Hamil}) immediately follows that the total spin $S^2$ is conserved and 
 ${\cal H}_I$ splits according the $su(2)$-irrep.
The multiplicity of each irrep associated with the total spin quantum number S is given by
\begin{equation}
n(S,N)=\frac{N!\,(2\,S+1)}{(N/2+S+1)!\,(N/2-S)!}.
\end{equation}
One finds the following decomposition in invariant subspaces
\begin{eqnarray}
{\cal H}_I &=& \bigoplus_{S=S_m(N,I)}^{N/2}\bigoplus_{r=1}^{n(S,N)}{\cal H}_I(S,r),\\ \nonumber
 {\cal H}_I(S,r) &=& \bigoplus_{S^z=-S}^{min\,(I-N/2,\,S)} |I,\,S,\,r,\,S^z\rangle\otimes {\cal H}_b(N_b(I,S^z)),
\label{Split} \end{eqnarray}
where $S_m(N,I)=\, max\,(N/2-I,s),$ ( $s=0$ for $N$ even, and $s=1/2$ otherwise), ${\cal H}_b(N)$ denotes
the eigenspace, in ${\cal H}_b$, of $N_b=\sum_k n_k$ corresponding to the eigenvalue $N,$ 
$N_b(I,S^z)=I-N/2-S^z,$  $|I,\,S,\,r,\,S^z\rangle$ is a simultaneous eigenvector
of ${\cal I},\,S^2,\,S^z$ associated respectively to the eigenvalues $I,\,S\,(S+1),\,S^z.$    
The reference state $|0\rangle,$ belongs to the subspace  $ 
\oplus _I{\cal H}_I(N/2),$ with maximal total spin eigenvalue $S=N/2:$ this subspace will be denoted
by ${\cal H}^{sym},$ and referred to as the {\sl symmetric } subspace.
If ${\cal H}^{sym}_I \equiv {\cal H}^{sym}\cap {\cal H}_I,$ one has
\begin{eqnarray}
{\cal H}^{sym}_I &=&
 \mbox{span} \{ (S^+)^n |\psi^{(I-n)}_{0,k} \rangle\,|\, n=0,\ldots,I\}\\ \nonumber 
\mbox{dim}({\cal H}_I^{sym})  &=& \sum_{l=0}^{{ {min}}(N,I)}
\pmatrix{I-l+N_b-1\cr N_b-1}.
\end{eqnarray}
The orthogonal complement of ${\cal H}^{sym}_I$ will be denoted by ${\cal H}^{A}_I.$
It is the direct sum of all the sectors with non-maximal $S^2$-eigenvalue.
Before ending  this section we notice that an additional term of the form
$H^\prime= S^z\sum_{k} w_{k} (b_k+b_k^\dagger),$ would destroy the $u(1)$ symmetry
generated by ${\cal I},$ but not the $su(2)$ structure.
Such a term, considered in \cite{ZARA}, does not correspond to an energy-exchange
but is a source of pure decoherence.
Since our analysis relies on the invariant decomposition 
${\cal H}=\oplus_I {\cal H}_I$
this term  has been  omitted here. 
\section{ $1$-Excitation Space}%%%%%%%%%%%%%%%%%%%%%%%%%%%%%%%%%%%%%%%%%%%%%%%%%%%%%%%
\begin{figure}[htb]
\begin{center}
\unitlength1cm
\begin{picture}(12,6)
\put(1,1){\epsfxsize=7cm\epsfbox{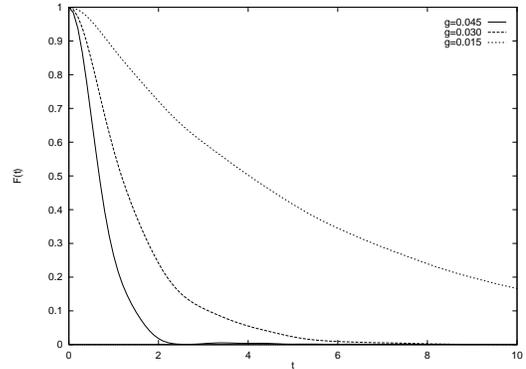}}
\end{picture}
\caption[]{ 
Fidelity as a function of time
for $|\psi_0\rangle=|\psi^{sym}\rangle.$ The coupling function
is $g_k(i)=g\; \forall k,i,$ ($N=2,\,N_b=200$).
The time unit is $\epsilon^{-1}.$
}
\label{Fig1}
\end{center}
\end{figure}

\begin{figure}[htb]
\begin{center}
\unitlength1cm
\begin{picture}(12,6)
\put(1,1){\epsfxsize=7cm\epsfbox{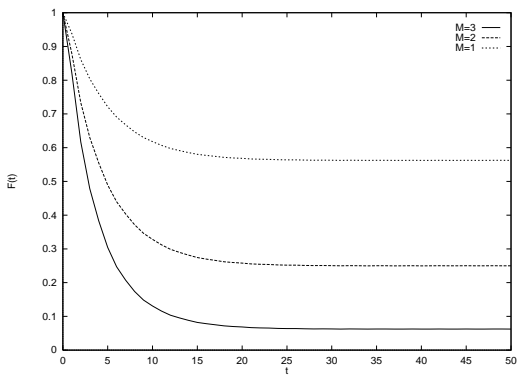}}
\end{picture}
\caption[]{ Fidelity as a  function of time for
$|\psi_0^s\rangle =M^{-1/2}\sum_{\alpha=1}^{M}|\alpha\rangle,\,(M=1,\,2,\,3;\,N=4,\,N_b=200).$ 
$g_k(i)=0.01\; \forall k,i.$
The time unit is $\epsilon^{-1}.$
}
\label{Fig2}
\end{center}
\end{figure}

\begin{figure}[htb]
\begin{center}
\unitlength1cm
\begin{picture}(12,6)
\put(1,1){\epsfxsize=7cm\epsfbox{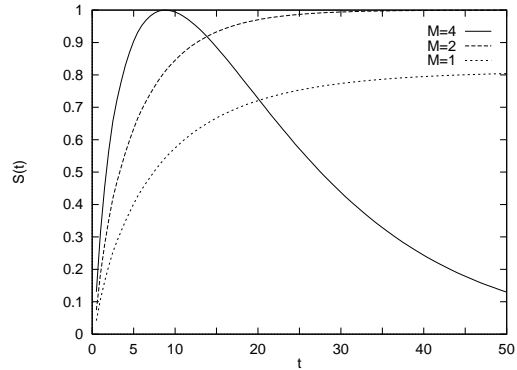}}
\end{picture}
\caption[]{ Entropy as a function of time for
$|\psi_0^s\rangle =M^{-1/2}\sum_{\alpha=1}^{M}|\alpha\rangle,\,(M=1,\,2,\,3;\,N=4,\,N_b=200).$
$g_k(i)=0.01\; \forall k,i.$
The time unit is $\epsilon^{-1}.$
}
\label{Fig3}
\end{center}
\end{figure}

Due to the field-theoretic nature of our model, 
the dimensionality formula (\ref{Dim}) clearly shows that for
increasing excitation number $I$ the problem of diagonalizing $H$ becomes 
rapidly intractable.
In particular, a finite-temperature analysis (arbitrary number of excitations)
is very difficult.
Nevertheless one of the interesting features of quantum noise is to be active
also a $T=0,$ thanks to vacuum fluctuations. 
This latter issue can be addressed by exact numerical means, without
an artificious truncation of the bosonic space,
by noticing that 
 the one-excitation space ${\cal H}^{(1)},$ has dimension  $d_1=N+N_b,$
that is only a linear function of the total number of degrees of freedom.
The basis $\{|\psi^{(1)}_{\alpha,k}\rangle\}$   is given by
$
%\begin{eqnarray}
|\alpha\rangle \equiv \sigma^+_\alpha\;|0\rangle,(\alpha=1,\ldots,N)$ and
$|k\rangle \equiv b^\dagger_k\;|0\rangle, (k=1,\ldots,N_b).\label{Basis} 
%\end{eqnarray}
$
Equation (\ref{Split}) in this case reads 
\begin{equation}
{\cal H}_1={\cal H}_1(N/2,1)\bigoplus_{r=1}^{N-1}{\cal H}_1(N/2-1, r),
\end{equation}
The symmetric space ($S=N/2$)  is  $N_b+ 1$-dimensional and it is  spanned by the vector
$|\psi^{sym}\rangle\equiv N^{-1/2} S^+\,|0\rangle,$ and by the whole set $\{|k\rangle\}.$
The  subspace ${\cal H}^A_1$ corresponds to $S=N/2-1.$
An orthonormal basis of ${\cal H}^A_1$ is given by 
\begin{equation}
|\phi_k\rangle\equiv S_k^+\,|0\rangle,\qquad 
S_k^+\equiv N^{-1/2} \sum_{j=1}^{N} e^{i\,k\,j} \sigma_j^+,
\label{eigen}
\end{equation}
where,
$k= 2\,n\pi/N,\,n=1,\ldots,N-1.$
Since in ${\cal H}^A_1$ the bosonic vacuum  factorizes, this subspace, when necessary,
will be identified with its projection over ${\cal H}_s$.
Now we observe that the vectors $\{|\phi_k\rangle\},$  are 
annihilated by $S^-,$ as they have minimal $S^z$-projection, but also  by the $\{b_k\}$,
as they have empty boson sector. From this follows that
$H_I \,|\phi_k\rangle =0,(\forall k) $ therefore ${\cal H}^A_1$ is {\sl decoupled from the bath};  
 it is an energy eigenspace with eigenvalue $E=\epsilon.$
In  terms of evolution superoperators, if $\rho$ is a state over ${\cal H}^A_1,$
 we have ithe fixed-point relations
${\cal E}_t^{0}(\rho)=\rho,\,(t\ge 0),
$
where ${\cal E}^0_t$ denotes the  superoperator associated with the bath-vacuum
density matrix $|0\rangle_b\langle 0|_b.$
The states over ${\cal H}^A_1$ {\sl are unaffected by the decoherence induced by  coupling
with the bath vacuum} and can therefore to encode information in a safe way.
The space ${\cal H}^A_1$ is noiseless only at  zero temperature;
for finite temperature the $|\phi_k\rangle$'s get mixed with all the vectors
belonging to the same $su(2)$-irrep, making the induced dynamics non unitary.
It is important to notice that, for $N\neq 2,$ this states are {\sl not} the
noiseless ones introduced in \cite{ZARA},
as the latter are associated with spin singlets (i.e. are annihilated by $S^-$ {\sl and} $S^+$)
and are decoherence-free at any temperature,
whereas the $|\phi_k\rangle$'s belong to  $N-1$-dimensional $su(2)$ multiplets.
The spectrum in the symmetric subspace  can be obtained by resorting to
exact numerical diagonalization of $H$, that provides the   
 eigenvectors and eigenvalues $\{ |\phi_i\rangle, E_i\}_{i=1}^{d_1}.$
\\
On the other hand the spectrum 
in ${\cal H}^{sym}_1$  is given by
 the $N_b+1$ zeros of the expression \cite{DAKO}
\begin{equation}
P_{N,N_b}(E)=E-\epsilon-N\,\sum_{k=1}^{N_b}\frac{|g_k|^2}{E-\omega_k},
\label{iter}
\end{equation}
that corresponds to the analogous single spin problem with rescaled coupling
$g_k\mapsto \sqrt{N}\,g_k.$
This follows from the symmetry constraint that makes $|\psi^{sym}\rangle$   
the only state  coupled with the bosonic modes.
Let $|\psi_0\rangle =\sum_i c_i^0\,|\phi_i\rangle,\;( c_i^0=\langle \phi_i|\psi_0\rangle)$
be the initial state; at $t>0$ we can  write, in terms of the chosen basis
\begin{eqnarray}
|\psi(t)\rangle &\equiv& e^{-i\,H\,t}|\psi_0\rangle=\sum_{\gamma=1}^{d_1}C_\gamma(t) \,|\gamma\rangle
\in {\cal H},\\
\nonumber  
C_\gamma(t) &= &\sum_{i=1}^{d_1}c^0_i\,c^i_\gamma e^{-i\,E_i\,t},\qquad (c^i_\gamma\equiv 
\langle \gamma|\phi_i\rangle).
\end{eqnarray}
The marginal density matrix is given by $\rho_s(t)=\mbox{tr}^b |\psi(t)\rangle\langle\psi(t)|$.
By using the relations
\begin{eqnarray}
\mbox{tr}^b |\alpha\rangle\langle\alpha^\prime| &=&\sigma^+_\alpha|0\rangle_s\langle 0|_s\sigma^-_{\alpha^\prime},\;
\mbox{tr}^b |k\rangle\langle k^\prime| =\delta_{k k^\prime} |0\rangle_s\langle 0|_s,\\ \nonumber
\mbox{tr}^b |\alpha\rangle\langle k|&=&  \mbox{tr}^b |k\rangle\langle \alpha|=0,
\end{eqnarray}
one obtains
\begin{eqnarray}
\rho_s(t) &=&\sum_{\alpha \alpha^\prime=1}^{N}C_\alpha(t)\,\bar C_{\alpha^\prime}(t)
 |\alpha\rangle\langle\alpha^\prime|\\ \nonumber &+&
|0\rangle_s\langle 0|_s  \sum_{k=1}^{N_b} |C_k(t)|^2.
\end{eqnarray}
The first (last) $N$ ($N_b$) terms describe a sector with a reversed (excited) spin (boson).
The marginal density matrix can be readily diagonalized, simply by observing
that it can be written in the form $\rho_s(t)= P_1(t)\,|\psi_s(t)\rangle\langle\psi_s(t)|+ P_0(t)
|0\rangle_s\langle 0|_s,$ where
\begin{equation}
|\psi_s(t)\rangle =\frac{1}{\sqrt{P_1(t)}}\sum_{\alpha=1}^{N}C_\alpha(t)\,|\alpha\rangle\in {\cal H}_s^{\otimes N},
\end{equation}
and
$
P_0(t)=1-P_1(t)=
\sum_{k=1}^{N_b} |C_k(t)|^2.
$
The von Neumann entropy of $S$ is therefore given by 
$
S_s(t)=-\sum_{i=0}^1P_i(t)\log_2 P_i(t).
$
A completely symmetric expression ($\alpha\leftrightarrow k,\, P_1\leftrightarrow P_0$) is obtained 
for the bath marginal density matrix $\rho_b=\mbox{tr}^s(\rho),$
from which follows that $S_b=S_s,$ furthermore
we have $S(S|B)=S(B|S)=-S_s$ for the conditional entropies, and $S(B:S)=-2\,S_s$ for the mutual entropy;
this is a consequence of the purity of the overall
system-bath state.\\
In order to study the corruption of the information stored in a  pure quantum state  it appears  useful
to study the following quantity,  called  (input-output) {\sl fidelity} \cite{SCH}
\begin{equation}
F(t)= \langle \psi_0^s|\,\rho_s(t)\,|\psi_0^s\rangle.
\label{Fidelity}
\end{equation}
The fidelity measures the overlap between the initial state $|\psi^s_0\rangle$
and the evolved one.
In the following we shall be interested in the evaluation of (\ref{Fidelity})
with initial data in ${\cal H}_1$ of the form $|\psi_0\rangle=|\psi_0^s\rangle\otimes |0\rangle_b,$
($ \rho_s(0)=|\psi_0^s\rangle\langle \psi_0^s|$). For such initial preparation, if 
$|\tilde \psi_s(t)\rangle= \sqrt{P_1(t)}\,|\psi_s(t)\rangle,$
one can write 
\begin{eqnarray}
F(t) &=& 
\langle\tilde \psi_s(t)|\,\rho_s(0)\,|\tilde \psi_s(t)\rangle\\ \nonumber
&=&|\langle \psi_0^s|\tilde \psi_s(t)\rangle|^2=
|\sum_{\alpha=1}^{N} C_\alpha(t) \,\bar C_\alpha(0)|^2\equiv|D(t)|^2.\label{Fid2}
\end{eqnarray}
The "decoherence" function $D(t)$ is related to the decay of the off-diagonal elements
of $\rho_s(t)$; indeed if, at $t=0$,  we prepare the system in the state 
$2^{-1/2}\,(|0\rangle_s+|\psi_0^s\rangle)\otimes |0\rangle_b,$
where $S^z\,|\psi_0^s\rangle= (1-N/2)\,|\psi_0^s\rangle,$
 it is immediate to check that
 $\langle \psi_0^s|\rho_s(t)|0\rangle_s=2^{-1}\,D(t).$
The mechanism  responsible for the energy exchange induces also
a dephasing between the zero and one excited spin states: dissipation is associated with decoherence.\\
In figure (\ref{Fig1}) is shown the behaviour of $F(t),$ for different values of the coupling with the bath,
with initial data $|\psi_0^s\rangle=|\psi^{sym}\rangle.$
In this figure and in the subsequent ones $\epsilon^{-1}$ is chosen as the time unit.
These results are obtained by exact diagonalization of $H$ in ${\cal H}_1,$ that provides
the dynamical functions $\{ C_\gamma(t)\}$. 
For strong  bath-system couplings $F(t)$ develops oscillatory structures, due to the back and forth
exchange of energy between the system and the bath.
We report the simulations for weak couplings, since it is the case physically relevant.
Furthermore, since we are essentialy interested in the role played   in the large times dynamics,
by the symmetry structure of the initial state, and not in a detailed description
of the bath-system coupling, we have choosen $g_k=g_0,\,(\forall k).$  
From the point of view of the energy-information loss
the latter choice is the worst case in that  each  qubit is coupled equally well with all the bath modes,
which should not, of course, be the case in real systems.
The fidelity in this range of coupling parameters and for intermediate times, vanishes in exponential way
$F(t)\simeq\exp(-t/\tau).$
The relaxation time $\tau,$ which turns out to be inversely proportional to $N\sum_k|g_k|^2,$
 is the time scale over which the dissipative process takes place.
The real and immaginary parts of $D(t)$ have an exponential damping  modulated by oscillations
over a time scale $\epsilon^{-1}.$
For very small times a naive perturbation up to the second order in $H_I$
shows  that indeed $F(t)\stackrel{t\rightarrow 0+}{\simeq}1-t^2/2\,\,N\,\Delta,$ 
where $\Delta=\sum_k |g_k|^2.$ 
Of course this process is nothing but the relaxation of the excited spin,
whose energy is transferred to the environment; for sufficiently large times one finds
\begin{equation}
U_t\,|\psi^{sym}\rangle\otimes|0\rangle_b=|0\rangle_b\otimes|\psi_b\rangle,
\end{equation}
where $|\psi_b\rangle=\sum_k c_k\, |k\rangle,$ is a superposition of all  one boson states.\\
Some words of caution are now in order.
The model under consideration is nothing but a  multi-mode generalization
of the Jaynes-Cummings model with many atoms. In analogy with  the latter,
for long time scale, $t > t_C,$ it exhibits a complex pattern of collapses and
revivals \cite{MON}. Furhermore since in each excitation space we have only a 
finite number of degrees of freeedom the phenomenon of the Poincar\'e recurrences 
is also present for $t> t_R.$
In the following we will show results for  $t\ll t_C,\,t_R,$
in other terms we assume that, thanks
to the great number of bosonic modes and the weak coupling,
 the physically relevant time-scales are
much smaller than the ones at which this more complex behaviour appear.
The energy exchange of the register with the bath can then  be considered
irreversible.
%%%%%%%%%%%%%%%%%%%%%%%%%%%%%%
%%%%%%%%%%%%%%%%%%%%%%%%%%%%%%
Suppose now that $|\psi_0^s\rangle= c_s\,|\psi^{sym}\rangle+ c_a\,|\psi^a\rangle,$ where
$|\psi^a\rangle\in{\cal H}_1^A,$ is a normalized vector, and $|c_a|^2+|c_s|^2=1.$
Also, for the sake of concreteness, and without any loss of generality, let us  consider the case $N=2.$
The vectors $|\psi^{sym}\rangle,\,|\psi^a\rangle,$ are given respectively by the two Bell-basis states
$
|\psi^{sym}\rangle = \frac{1}{\sqrt{2}}( |\uparrow\downarrow\rangle+|\downarrow\uparrow\rangle),$
$|\psi^a\rangle = \frac{1}{\sqrt{2}}( |\uparrow\downarrow\rangle-|\downarrow\uparrow\rangle).
$
The initial marginal density matrix of $S$ is therefore given by
\begin{eqnarray}
\rho_s(0)&=& |c_s|^2\,|\psi^{sym}\rangle\langle\psi^{sym}|+|c_a|^2\,|\psi^a\rangle\langle\psi^a|
\\ \nonumber
&+& c_s\,\bar c_a\,|\psi^{sym}\rangle \langle\psi^a|+ \bar c_s\, c_a\,
|\psi^\alpha\rangle \langle\psi^{sym}|,
\end{eqnarray}
By using our  previous result for the symmetric initial state, and the fact that $|\psi^a\rangle$
is an energy eigenstate, it follows easily from (\ref{Fid2}),  for $t$ large enough,  that
\begin{eqnarray}
{\cal E}^0_t(|\psi^a\rangle\langle\psi^{sym}|)&=&{\cal E}^0_t(|\psi^{sym}\rangle\langle\psi^a|)=0, \\ \nonumber 
{\cal E}^0_t(|\psi^a\rangle\langle\psi^a|)&=&|\psi^a\rangle\langle\psi^a|,\; 
{\cal E}^0_t(|\psi^{sym}\rangle\langle\psi^{sym}|)=|0\rangle_s\langle 0|_s.
\end{eqnarray}
Therefore 
the large times density matrix is given by
\begin{equation}
\rho_s\simeq |c_a|^2 |\psi^a\rangle\langle\psi^a| + |c_s|^2 |0\rangle_s\langle 0|_s,
\label{Rout}
\end{equation} 
from which straightforwardly follows  for the fidelity the  behaviour
\begin{equation}
F{\simeq}|c_a|^4=(1-|c_s|^2)^2=(1-|\langle\psi^{sym}|\psi_0^s\rangle|^2)^2.
\label{Fout}
\end{equation}
In other terms:
 the final state  depends on  the initial preparation symmetry content;
 a complete corruption of the initial information 
 is obtained only if the initial state belongs to the vacuum  $S^2$-sector ${\cal H}_1(N/2)$,
so that the smaller is  the projection over it, the closer  to one is the fidelity.
\\
The extreme case is $|\psi_0^s\rangle\in{\cal H}_1^A,$
$F(t)=1,\,\forall t$ in which  there is no relaxation at all.
In the intermediate situations the spin system remains partially entangled with the
environment and its state never gets pure.
%%%%%%%%%%%%%%%%%%%%%%%%%%%%%%
This situation is illustrated in figures (\ref{Fig2}), and 
(\ref{Fig3}) where  fidelity and entropy are shown as functions of time 
in the case of  $|\psi_0^s\rangle =M^{-1/2}\sum_{\alpha=1}^{M}|\alpha\rangle,$ for $M=1,\,2,\,3.$
In this case  it is trivial to check, by using equations (\ref{Rout}), and (\ref{Fout}),
  that  $F\simeq(1-M/N)^2,$  and $S_s
\simeq (M/N-1)\log_2 (1-M/N)-M/N\log_2(M/N).$
Notice that if $|c_s|^2=1,$ one has the complete de-excitation of the spin system, therefore
the initial state $|\psi^{sym}\rangle$   is maximally entangled
and the  final state $|0\rangle_s,$
with zero mutual entanglement of the qubits.
The system undergoes  energy as well as {\sl information} loss.   
 On the other hand 
 if $|c_a|^2=1,$ the final, and initial, state $|\psi_a\rangle$
is  maximally entangled: energy and information are conserved. 
%%%%%%%%%%%%%%%%%%%
%%%%%%%%%%%%%%%%%%%%
In the temporal range in which the decay of fidelity is exponential,
the following relation between relaxation times holds:  $\tau(c_s)\,|c_s|^2 =\tau(1).$
%%%%%%%%%%%%%%%%%%%
%%%%%%%%%%%%%%%%%%%%
\section{Replica dependent coupling}
\begin{figure}[htb]
\begin{center}
\unitlength1cm
\begin{picture}(12,6)
\put(1,1){\epsfxsize=7cm\epsfbox{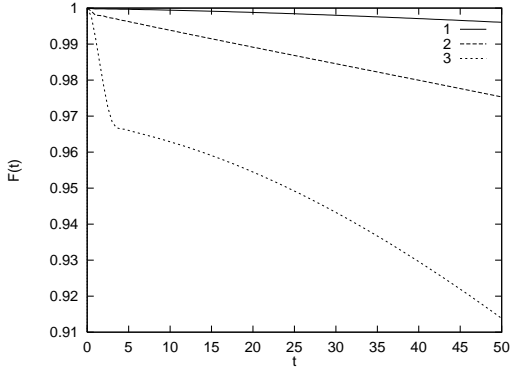}}
\end{picture}
\caption[]{ Fidelity as a function of time for
$|\psi_0^s\rangle\in{\cal H}_1^{A},$  with $\xi=10\, (1),\,\xi=5\, (2),\,\xi=1\, (3).$
The coupling function
is $g_k(i)=g_0\,\cos(k\,i/\xi).$
$N=2,\,N_b=200,\,g_0=0.01.$
The time unit is $\epsilon^{-1}.$
}
\label{Fig4}
\end{center}
\end{figure}

\begin{figure}[htb]
\begin{center}
\unitlength1cm
\begin{picture}(12,6)
\put(1,1){\epsfxsize=7cm\epsfbox{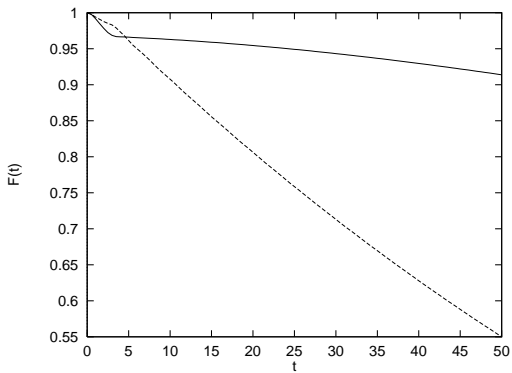}}
\end{picture}
\caption[]{
Fidelity as function of time for
$|\psi_0\rangle=|\psi^{sym}\rangle,$ (dashed line)
and $|\psi_0^s\rangle\in{\cal H}_1^{A},$ (solid line).
The coupling function
is $g_k(i)=g_0\,\cos(k\,i).$
$N=2,\,N_b=200,\,g_0=0.01.$
The time unit is $\epsilon^{-1}.$
}
\label{Fig5}
\end{center}
\end{figure}

In this section we present some numerical results in the case in which the system-bath coupling depends on the 
qubit replica. 
If the coupling functions $\{g_k(i)\}$ and/or the qubit energies $\{\epsilon_i\}$ depend on 
the qubit replica, the total spin operator
$S^2$ is no longer a constant of the motion.
This situation is expected to be more realistic than the one previously assumed
in that the latter amounts to have a bath with an infinite (i.e. very large) coherence length.
In this case the decomposition (\ref{Split}) is not invariant:  the 
dynamics results in a non-trivial mixing of the $su(2)$-irreducible sectors ${\cal H}_I(S,r).$
In particular one has $S^-{\cal H}_1^A\neq 0,$ therefore the vectors $|\psi^a\rangle$ can decay. 
In other words the loss of the symmetry constraint allows the dissipation-decoherence
induced by the bath to invade the whole Hilbert space. 
We choose
%\begin{equation}
$g_k(i)=g_0\,\cos(k\,i/\xi),$
%\end{equation}
where $\xi$ is a parameter related to the bath coherence length 
(so that for $\xi=\infty$ we recover the results of the previous
 sections).
In figure (\ref{Fig4}) are reported the plots of $F(t)$ with initial condition in ${\cal H}_1^A$
for different $\xi$'s.
%%%%%%%%%%%%%%%%%%%
%%%%%%%%%%%%%%%%%
Figure (\ref{Fig5})  shows the behaviour of $F(t)$ for small times with $\xi=1.$
Notably one observes that the initial condition  $|\psi_0^s\rangle\in{\cal H}_1^{A},$
exhibits  a faster fidelity decay with respect to $|\psi_0^s\rangle\in{\cal H}_1^{sym},$ 
for short times $t<t_c$. 
For longer times, with  obvious meaning of the notation,  $F_A(t) > F_{sym} (t).$ 
The numerical simulations show in any case that $\bar F_A> \bar F_{sym},$ the bar denoting  temporal
average. 
%%%%%%%%%%%%%%%%%%%
%%%%%%%%%%%%%%%%%%%%%%%%%
\section{conclusions}
In this paper we have presented the study of a physical model  for a 
quantum register coupled with the environment.
Information is encoded in the quantum state of the register.
The register consists of $N$ non-interacting replicas of a two-level system (i.e. a $N$-qubits register).
The register environment is described by a bosonic bath  consisting  of $N_b$ modes,
with $N_b\gg N.$
Its coupling with the register is realized by the exchange of elementary {\sl quanta}
of energy. The resulting dynamics of the register is dissipative;
in the weak coupling regime,  energy and information  are irreversibly lost into the bath.
Even though the model is non-trivial, exact 
analytical as well as numerical results can be obtained thanks to the existence
of a constant of the motion (excitation number) that leads to a decomposition of the total Hilbert
space in dynamically independent sectors.
Assuming that the environment couples  in the same way with all the register qubits
one has a further splitting of the sectors according to the irreducible representations of the 
spin $su(2)$ algebra. Each $su(2)$ lowest vector is decoupled from the  bath vacuum fluctuations
and therefore, at zero temperature, is decoherence free.  
The smallest subspace in which one can have non-trivial physics is the one-excitation
sector ${\cal H}_1.$
The  dimension of  ${\cal H}_1$ scales linearly in the total number of degrees of freedom,
therefore a thorough analysis of the dynamics in ${\cal H}_1,$ can be performed
by means of exact numerical diagonalization of the model Hamiltonian.
The temporal dependence of quantity of interest, such  as  fidelity and entropy have been studied.
The  asymptotic behaviour depends on the symmetry content of the initial state.
Smaller is the projection of the initial state over the vacuum $su(2)$-sector ${\cal H}^{sym},$ 
greater is the fidelity. 
In particular a complete energy-information loss
 occurs only when the state
belongs to ${\cal H}^{sym}.$
Some numerical results for bath-system coupling dependent on the qubit are also presented.
In this more realistic situation the $su(2)$-structure is unstable:
dissipation and decoherence affects the whole Hilbert space and then 
safe encondings no longer exists.
Nevertheless our results
 shows that, on long time scales, the average fidelity of the previously noiseless states
is still greater of the one of the other states.
This suggest that the symmetry-based protection of quantum state suggested in \cite{ZARA}
can be valuable in the general case.
This last issue, along with the necessary 
finite temperature generalizations, worth further investigations.\\
\begin{acknowledgments}
Stimulating discussions with M. Rasetti and R. Zecchina
are gratefully aknowledged.
The author also thanks  
C. Calandra and G. Santoro for providing him access to the CICAIA of the Modena University,
and  Elsag-Bailey for financial support
\end{acknowledgments}

%%%%%%%%%%%%%%%%%%%
\end{multicols}
\end{document}